# Millimeter-wave in-package: tackling the system-in-package interconnection paradigm

Souradip Sarkar  
Nokia Bell Labs

Gert-Jan Stockman  
Nokia Bell Labs

Brecht Francois  
Nokia Bell Labs

*Abstract*—This work describes the architecture and implementation of a high-data-rate, energy-efficient millimeter-wave (mm-wave) waveguide solution for integration inside an Integrated Circuit (IC) package. The complete waveguide solution (together with couplers) has been designed to fit into the tight dimensional requirements of a wafer bonded 3D package and provides an extremely wide bandwidth in the mm-wave spectrum. To achieve this, we investigate the use of new high-permittivity materials. To be able to fulfill the data-rate demands of modern System-in-Packages (SiPs), Monolithic Microwave Integrated Circuits (MMICs) and integrated high performance analog ICs together with Digital Signal Processing (DSP), our solution is inherently scalable and adaptable in several ways. The work is novel and also discusses the various opportunities and challenges associated with the solution.

## I. Introduction

The ongoing increase in processing power of Multi-processor System-on-Chip (MPSoC) Integrated Circuits (ICs), is being fueled by advanced integration, scaling, improvements in process technology, advancements in packaging technologies, and the shift in the fundamentals of computer architecture from single- to multi-core platforms. GPUs and other parallel multi-core devices are being used extensively in a wide variety of applications (from data-centers, personal computers, mobile handhelds and self driving cars to autonomous robots and even network equipment). The bandwidth requirement inside a modern processing module or chip is of the order of several Tb/s range, and this trend seems to be steadily growing. Modern Multi-Chip Modules (MCMs) comprise of several ICs packaged together in a common housing (Flip-Chip Ball Grid Array (FC-BGA) or Plastic Ball Grid Array (P-BGA)). The total off-chip bandwidth can range somewhere between $5 - 20$ Tb/s depending on the application, and this demand is set to grow further with the integration of complex signal processing MPSoCs, GPUs, and MCMs all in the same package. As the physical size of package pins has reached its lower-limit (a reduction in size increases resistivity which in turn decreases the speed of the link), the number of package pins can not grow at the same rate at which the demand for off-chip data bandwidth is growing. Serial link bandwidth is still growing steadily, however, the share of communication energy cannot increase in a similar trend as systems are becoming power limited. Consequently, improvements in energy dissipation have been indispensable and its evolution is shown in Fig. 1. This graph clearly illustrates how serial links have ad-vanced over the years. The total energy is normalized with respect to the supported bandwidth, to highlight the link energy improvements. Over the last two decades the link energy has reduced from over $100\,\text{pJ/bit}$ to less than $1\,\text{pJ/bit}$. Different IPs, MPSoCs and DRAMs all communicate to one another using these electrical I/Os (together with their transceivers they are referred to as SerDes). SerDes designs have only advanced approximately at a modest $33\%$ per year (roughly $10\times$ over the last $8$ years) but typically, standards like PCIe, QPI, DDR4 and HDMI are lagging behind research and move at an even slower pace. In order to support the I/O and memory bandwidth demands of complex System-on-Chips (SoCs) and top of the line CPUs and GPUs, several SerDes links are used in parallel. The share of the communication power, especially in SerDes and DRAM physical layer (PHY) power as a fraction of the total power budget can be seen in [1] and [2]. It approximately accounts for about $20 - 30\%$ of the total system energy budget. By extension, off-chip and on-board communication is becoming a serious design problem with integration of multiple high bandwidth IC packages on the same board.

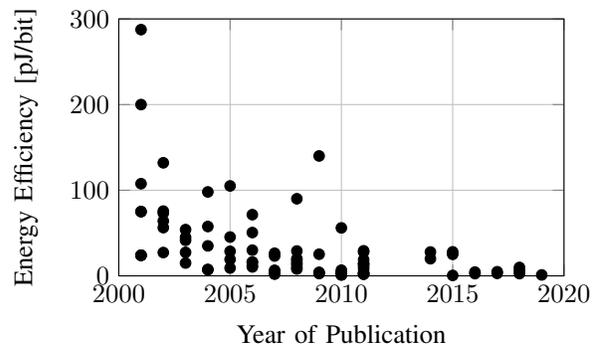

Fig. 1. Evolution of the energy efficiency in high-data-rate SerDes research. Courtesy: ISSCC conference proceedings.

In addition to digital MPSoCs, mixed signal RF-ICs and SoCs often face yet another challenge: handling extremely high frequency signals (in the $30-300\,\text{GHz}$ millimeter-wave (mm-wave) spectrum). From package pins and bond wires to termination circuits, they all demand cautious planning and design. Getting the extremely high frequencies into and out of the analog front-end (Analog to Digital convertor (ADC) or Digital to Analog convertor (DAC)), typically incurs a significant loss for high frequencies over traditional copper interconnects, because of the inherent low-pass channel characteristics of resistive interconnects. Henceforth, we have explored the feasibility of deploying a Dielectric Rod Waveguide (DRW) at the package level (System-in-Package (SiP)), and the challenges associated with it. Our work is focused on bringing the mm-wave processing inside IC packaging (leading to further miniaturization), but shall also be relevant to RF-SoCs (primarily catering to RF signal processing for 5G (mm-wave spectrum) communications along with bandwidth hungry complex signal processing ICs in next-generation test and measurement equipment like oscilloscopes, network and spectrum analyzers.

The rest of the paper is organized as follows: Section II presents the prior research performed in the domain. Section III presents detailed background on the proposed dielectric waveguide and the other relevant elements of the system. The 3D electromagnetic (EM) simulation for the different scenarios have been addressed in Section IV, and finally we conclude in Section V.

## II. Related Work

The first non-radiating dielectric waveguide for mm-wave applications was presented by Yoneyama and Nishida in [3]. Thereafter, it has been widely adopted in several applications in the mm-wave domain. It is seen as a perfect candidate for Monolithic Microwave Integrated Circuits (MMICs) and for 5G mm-wave applications. Dielectric waveguides for board level communication have been presented in [4, 5, 6, 7]. Even though the work in [5] already mentions chip-to-chip communication, the cross-sectional dimensions of the proposed waveguide are in the range of $2\,\text{mm} \times 1\,\text{mm}$, making it impractical for integration inside a package. The work in [8] is the first demonstration of a Rogers TMM3 ceramic composite material based waveguide (dielectric constant of $\epsilon_r = 3.27$) integrated in an IC together with the transceivers. The dimensions of the waveguide are around $400-500\,\mu\text{m}$ (which is large compared to the size of modern packages) and achieves a pass-band frequency range of $220-270\,\text{GHz}$. This is the first confirmed result that manages to bring waveguide technology to SoCs and inside packages with low attenuation of the wave propagating through it. Chip-to-chip interconnects are still mostly copper based electrical links either in a 2D or 3D setting. Typically, for high bandwidth demands, PCIe is the board level interconnect standard, where the PHY uses multiple SerDes lanes (each lane typically includes differential copper traces). We have not comprehensively compared our solution to the state of the art SerDes links, as they are complementary in terms of technology. Research on photonic interconnects inter- and intra-chip has been reported in [9]. Photonic interconnects offer very wide bandwidths. The challenge however is the Electrical-to-Optical (EO) and Optical-to-Electrical (OE) conversion. Wireless interconnects too have been considered, not just for chip-to-chip [10] but also within the same chip [11]. However, as these wireless solutions are bound by the Free Space Path Loss (FSPL) equation (loss increase quadratic with distance), they typically suffer more from attenuation for short distances in the mm-wave band compared to the guided loss in a non-radiating dielectric waveguide.

## III. Proposed Waveguide solution

This section introduces the overall system architecture, together with a detailed explanation of the essential constituent elements. A theoretical foundation of the propagating EM modes in this guided dielectric medium is also discussed.

### A. System architecture

The schematic representation of the envisioned system is presented in Fig. 2. At the transmitting end, the signal of interest is up-converted from baseband into the mm-wave spectrum, whereafter it is coupled into the dielectric waveguide. The EM wave propagation happens in the guided dielectric domain. At the receiving end, an identical coupler is employed to convert the EM energy from the guided domain back to the electrical domain, which is then in turn down-converted back to baseband. For certain applications that are envisioned to operate within the mm-wave band, such as in 5G, the up- and down-conversions can be omitted. To illustrate the coupling from electrical to guided domain, we present the fields propagating through a simple adiabatic coupler in a cross-section along the direction of propagation in Fig. 3. Clearly, the electrical mode (left side) is converted from the rectangular metallic waveguide to the rectangular dielectric waveguide (middle) and back to the electrical mode (right side) without visual mode conversion when comparing output to input. Observing the scattering parameters of this solution yields a virtually lossless propagation (no loss tangent considered yet).

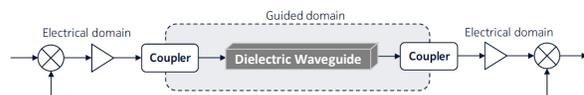

Fig. 2. A schematic representation of the proposed waveguide solution. Up- and down-convertors allocate the signals to a suitable frequency band, while couplers take care of the conversion from the electrical to the guided dielectric domain, and back.

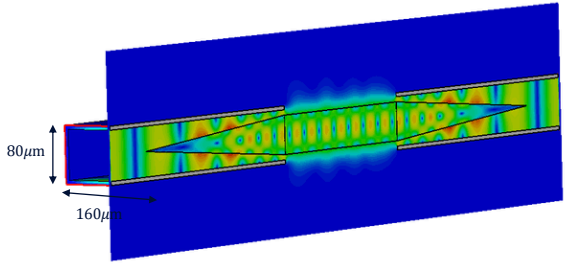

Fig. 3. Magnitude of the electric field vector throughout a rectangular DRW ($\epsilon_r = 1000$, $a \times b = 160\,\mu\text{m} \times 80\,\mu\text{m}$) together with a simple adiabatic coupler design illustrating the conversion and conservation of the propagating mode between input and output.

### B. Dielectric Waveguide

Dielectric waveguides can be composed of a large variety of materials, shapes and dimensions. Our waveguide is a DRW, a long, rectangular in cross-section, solid piece of extremely high dielectric material (say relative permittivity, $300 < \epsilon_{r_1} < 7000$) as shown in Fig. 4. The advantage of using a high permittivity material is

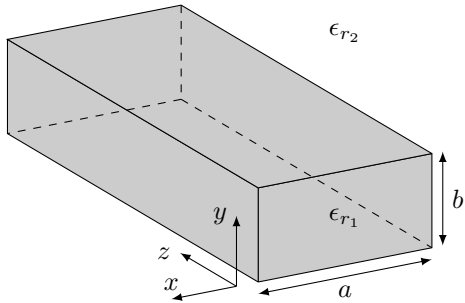

Fig. 4. Artist's impression of a rectangular DRW with cross-sectional dimensions of $a \times b$ and arbitrary length.

the reduction of the waveguide's dimensions making it potentially fit inside an IC package while keeping the lower cut-off frequency in the mm-wave band. The proposed solution can transition to standard rectangular waveguides at the board level (on-board, and in-board variants [5] and [6]) making use of a custom-designed coupler. Advancements in material processing and manufacturing components have enabled extreme precision in fabrication of the required dimensions. Furthermore, predictions hint that a low loss tangent for high-dielectric materials could potentially be achieved given sufficient traction for these materials pushes manufacturers' focus more towards low-loss [4]. This work will focus on a simulation-based feasibility analysis of novel extreme high dielectric constant materials like Barium Titanate, Lead Zirconate Titanate, Strontium Titanate [12], Barium Strontium Titanate or conjugate polymers [13]. Their permittivities enable the waveguide to shrink to less than $100\,\mu$m in height, which is essential for integrating it in a modern 3D Stacked Integrated Circuit (SIC) package [14]. An example is shown in Fig. 5 where five separate layers of a stacked die setup are shown. 3D

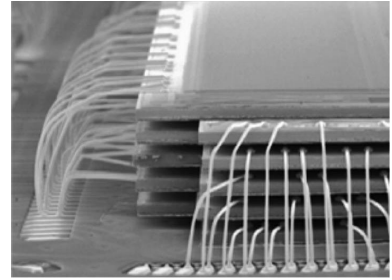

Fig. 5. Wire bonded chips in 3D SIC package. Courtesy: [14]

SICs have multiple SoCs or memory dies, stacked on top of each other in either Cu-Cu bonding based (face-to-face (F2F)), or TSVs/RDL/$\mu$bump based (face-to-back (F2B)) [14]. The waveguide is intended to confine the signal (that can typically interfere with neighboring wires) and transmit with low loss over the whole SiP. As shown in [5], the achieved guided attenuation is much smaller than what could be expected for a wireless solution (dictated by the FSPL equation).

In 1969, Marcatili, proposed a set of approximate equations for EM wave propagation through dielectric waveguides [15]. The primary assumption was that most of the signal power is confined and flowing inside the waveguide. This method was used as a first order approximation for calculating the propagation constant of a rectangular waveguide. The next step is deriving the EM mode profile in the medium. Various modes of propagation can be achieved depending on the geometry, dielectric constant of the material and the frequency of interest. The main advantage of the dielectric channel is the non-radiating behavior outside of the dielectric rod. Compared to wireless links, a waveguide achieves excellent signal confinement and isolation. Therefore, it could be extended to different stacked die layers, with fairly low attenuation. A low-loss channel is critical in limiting the total communication power. Otherwise, higher transmit power and channel equalizers are needed which typically are power hungry. In principle, multiple such waveguides can coexist and together they can provide extremely high bandwidth at the package level.

## C. Modes of Propagation

In a non-radiating DRW, the preferred propagation mode is either the hybrid Longitudinal Section Magnetic (LSM) mode or Longitudinal Section Electric (LSE) mode [4]. These modes are a subset of the hybrid EM modes in the sense that they have an electric and magnetic field component in the direction of propagation. Additionally, LSE and LSM modes have a component of either electric or magnetic field, respectively, that is zero in one of the transverse directions. Derivation of these modes is analogous to the derivation of transverse modes (TE and TM) where Maxwell's equations are solved under the assumption of the absence of electric or magnetic field in the direction of propagation. For LSE and LSM modes, the exact same approach can be used, with the difference that now one of the transverse components is chosen to be zero instead of the longitudinal component [16]. Our solution has LSE characteristics (Table I). The first three propagation modes were numerically calculated using CST. The magnitude of the electric field vector for these modes is shown in Fig. 6.

|  | $e_z = 0$ | $e_z \neq 0$ |
|---|---|---|
| $h_z = 0$ | TEM | TM |
| $h_z \neq 0$ | TE | LSE ($e_x = 0 \vee e_y = 0$) <br> HEM ⟨ <br> LSM ($h_x = 0 \vee h_y = 0$) |

TABLE I
CLASSIFICATION OF ELECTROMAGNETIC MODES BASED ON THE EXISTENCE OR ABSENCE OF CERTAIN FIELD COMPONENTS.

## D. Coupler

When connecting an Non-Radiative Dielectric (NRD) waveguide to a metallic waveguide, an appropriate mode converter is required (also referred to as the coupler). An ideal coupler performs both well in return and insertion loss for the in-band spectrum. Our mode converter, as shown in Fig. 3 is a simple tapered coupler that adiabatically couples the wave from the electrical domain to the guided domain. The geometry of the structure is important from a fabrication perspective, and the structure was chosen to maximize the degree of confidence for reliable operation.

## IV. SIMULATIONS

For the 3D EM simulation setup, we have considered a rectangular waveguide with $\epsilon_{r_1} = 1000$ and cross-sectional dimensions $80\,\mu$m $\times\, 160\,\mu$m. The relative permittivity of the waveguide surrounding was chosen as $\epsilon_{r_2} = 12$. It is assumed that either GaAs, InP or Si would be considered as surrounding material, which have relative permittivities (static) of 12.9, 12.5 and 11.7, respectively. However, at higher frequencies, the permittivity tends to be slightly lower. The propagation characteristics of the waveguide were simulated in CST Microwave Studio.

The couplers are adiabatic as shown in Fig. 3. The two couplers are connected to the two ends of the waveguide and mode conversion is achieved. We performed three separate set of experiments to better understand the challenge of deploying and studying characteristics of a realistic waveguide segment structures inside an IC package. The first setup was just the rectangular waveguide together with the couplers at both ends. The next scenario is aimed at characterizing bending loss in the waveguide, as it is evident that interconnects within a package are likely to have a few bends. Finally, crosstalk behavior has been simulated between waveguide structures running in parallel, to understand how close and how many such waveguides can be integrated in a package. This first scenario is the most simple yet complete representation of the entire channel. The channel length was considered to be 3 cm, matching the ultra short-reach package level dimensions. The insertion loss characteristic of the channel is shown in Table II.

|  |  | $f$ [GHz] | | | |
|---|---|---|---|---|---|
|  |  | 90 | 110 | 130 | 150 |
| $\tan \delta$ | 0 | 0.2 | 0.3 | 1.0 | 1.8 |
|  | 0.005 | 0.2 | 0.3 | 1.0 | 1.8 |
|  | 0.02 | 0.5 | 0.8 | 1.7 | 2.3 |

TABLE II
FREQUENCY DEPENDENT LOSS (DB/MM) OF THE DIELECTRIC MEDIUM FOR VARIOUS VALUES OF $\tan \delta$.

We have performed different simulations, one ideal case without loss tangent ($\tan \delta = 0$), the second with a realistic loss tangent ($\tan \delta = 0.0005$), and a worst case scenario ($\tan \delta = 0.002$) based on [12]. The channel shows a consistent return loss of 20 dB over most of the bandwidth. The loss in the forward transmission path is typically less than 2 dB/mm for the waveguide over the whole frequency spectrum.

Next, we performed a sweep analysis on the bending radius, to assess the impact of bending loss.

## A. Bending loss

The physical floor-plan inside a stacked 3D IC is challenging and a waveguide is expected to consist of multiple bends. Henceforth, a $\pi/2$ radians bending loss simulation for different bend radii $R$ was performed and shown in Fig. 7. In an effort to capture the additional loss introduced by a bend, a straight section, identical in length ($\pi/2 \times R$) is also simulated and subtracted from the loss in the bend. The bending loss is reported with respect to variation in the bend radius. The choice of a poor loss tangent ($\tan \delta = 0.002$) is deliberate to depict a realistic scenario as of today. We are investigating very sharp bends (that introduce large mode conversion) needed in an IC package environment. It is evident that the extremely high dielectric constant achieves a very good confinement already and the increasing confinement for increasing frequencies (and with it a decrease in bending loss) that we typically expect, is irrelevant in this case. The bending loss is dominated by mode conversion, as is apparent from the simulation of the modal distribution shown in Fig. 8. The signal enters the bend (bottom) in the fundamental mode but leaves the structure as a superposition of modes (right) due to conversion in the bend.

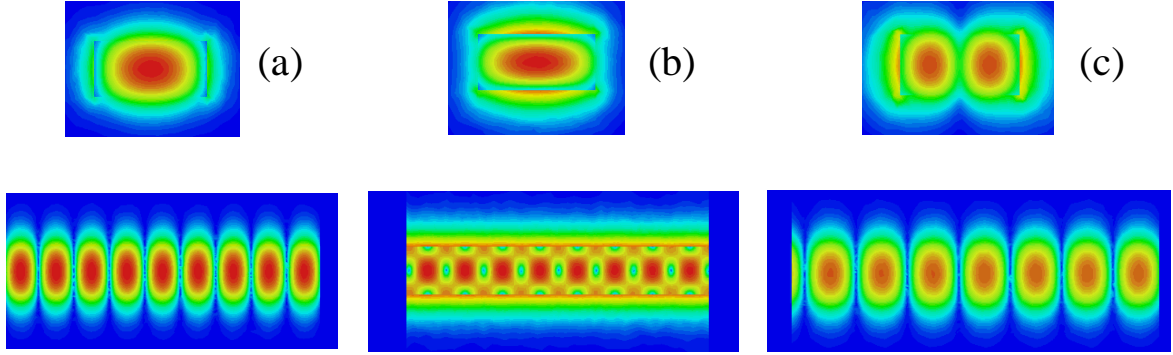

Fig. 6. Cross-sectional (top) and longitudinal (bottom) view of the magnitude of the electric field vector of three different propagation modes achieved.

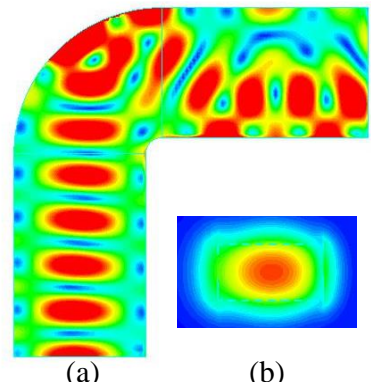

Fig. 8. The magnitude of the electric field along the waveguide at 100 GHz; (a) electric field vector along bending radius of 100 $\mu$m; (b) electric field pattern along the cross-section.

energy coupling was insignificant and it increases as the two waveguides are brought closer. This behavior is expected, thereby leading to an optimal or minimum separation between two parallel waveguide structures. More than 55 dB of loss over the entire spectrum, makes this solution extremely attractive from a signal integrity perspective.

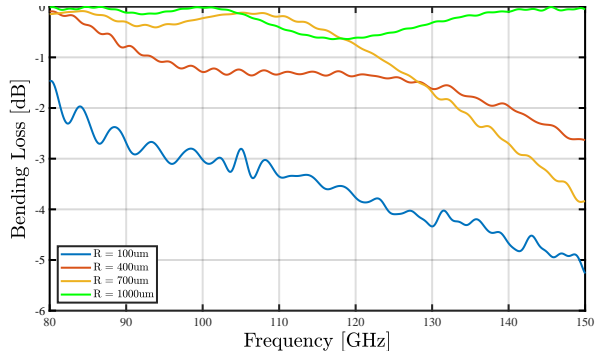

Fig. 7. The bending loss with respect to the bending radius over the relevant spectrum. Setting: $\tan \delta = 0.002$.

## B. Crosstalk

To model crosstalk, the simulation setup included two parallel waveguides (160 $\mu$m × 80 $\mu$m × 1 mm), where the distance between their outer edges ($d$) was varied. The near-end and far-end crosstalk over the entire spectrum are shown in Fig. 9 and Fig. 10, respectively. The loss tangent was excluded in this setup, so that purely the crosstalk effects of the dielectric waveguide could be captured. Excellent crosstalk behavior confirmed the expectations, as the extremely high dielectric constant confines the energy well. There is no need for shielding the waveguides, which can potentially introduce fabrication challenges. The amount of electromagnetic

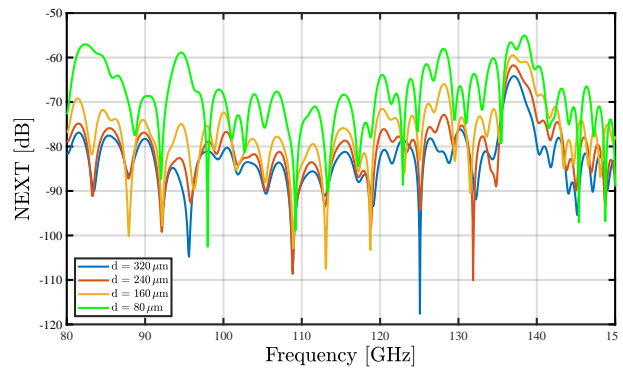

Fig. 9. The near-end crosstalk. The distance between the waveguides is represented by $d$.

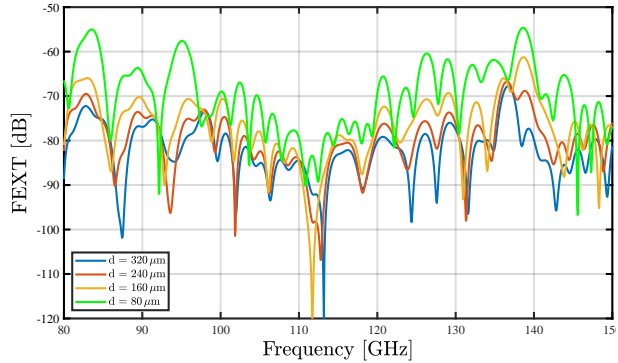

Fig. 10. The far-end crosstalk. The distance between the waveguides is represented by $d$.

## V. Conclusion and Future work

It is clear from the first hand simulation results that our proposal about having waveguides in a 3D SIC package offers a massive increase in channel bandwidth or data rate handling capacity within an IC package. The high dielectric constant offers very good confinement, and even though the dielectric losses are higher, the interconnect lengths inside a package are relatively small. The major challenge in integrating extremely high dielectric materials is their dispersion and dielectric loss tangent. Since, the materials under investigation have not been optimized, fabricated and measured in this particular deployment scenario, the loss tangent is not well characterized. Three different scenarios of loss tangent were considered for simulation purposes, disregarding the variation over the whole frequency range. With respect to dispersion, there is no literature on these extremely high dielectric constant materials. With respect to couplers, the mode converter presented can launch only the first mode. This motivates further research of an improved mode converter to simultaneously launch multiple modes, which can substantially increase the channel capacity. With respect to bending loss, it becomes apparent that as long as extremely sharp bends are avoided (less than $100\,\mu$m), the loss is insignificant. Another advantage of high dielectric materials is reduction of antenna effect losses. In addition, the crosstalk behavior is favorable, because of the high degree of confinement. This motivates further research in high dielectric materials together with reducing the loss characteristics.

To conclude, the research focuses on simulation based material innovation and presents the opportunity of significantly reducing power in the driver circuits and equalization techniques. Low loss characteristics opens up new opportunities in shared channel interconnects, advanced transceiver architectures, with higher order modulation and multiple SerDes channels mapped to a single waveguide. In addition, it is an effort to bring RF signal processing inside a complex SoC. However, the challenge in material level research, design of ideal mode converters together with reliable manufacturing is indispensable for commercial success of this solution.